\documentclass[conference,a4paper]{IEEEtran}

\usepackage{cmap}
\usepackage[utf8]{inputenc}
\usepackage[english]{babel}

\usepackage{indentfirst}
\usepackage{amsmath}
\usepackage{amssymb}

\usepackage{graphicx}
\usepackage{cite}
\usepackage[hyphens]{url}
\usepackage{subfig}
\usepackage{tikz}
\usepackage{listings}
\usepackage{hyperref}
\usepackage{siunitx}
\usetikzlibrary{arrows}
\usetikzlibrary{patterns}

\IEEEoverridecommandlockouts

\newcommand\copyrighttext{%
\footnotesize \textcopyright \enspace 2020 IEEE. Personal use of this material is permitted. Permission from IEEE must be obtained for all other uses, in any current or future media, including reprinting/republishing this material for advertising or promotional purposes, creating new collective works, for resale or redistribution to servers or lists, or reuse of any copyrighted component of this work in other works. DOI: \href{https://doi.org/10.1109/BlackSeaCom48709.2020.9234994}{10.1109/BlackSeaCom48709.2020.9234994}
}
\newcommand\copyrightnotice{%
\begin{tikzpicture}[remember picture,overlay]
\node[anchor=south] at (current page.south) {\fbox{\parbox{\dimexpr\textwidth-\fboxsep-\fboxrule\relax}{\copyrighttext}}};
\end{tikzpicture}%
}

\begin{document}
	
\title{Resource Allocation Strategies\\for Real-Time Applications in Wi-Fi 7\thanks{The work was carried out at NRU HSE and supported by the Russian Science Foundation (agreement 18-19-00580).}}

\author{\IEEEauthorblockN{
		\IEEEauthorrefmark{1}\IEEEauthorrefmark{2}\IEEEauthorrefmark{3}Evgeny~Avdotin,
		\IEEEauthorrefmark{1}\IEEEauthorrefmark{2}\IEEEauthorrefmark{3}Dmitry~Bankov,
		\IEEEauthorrefmark{1}\IEEEauthorrefmark{2}\IEEEauthorrefmark{3}Evgeny~Khorov,
		\IEEEauthorrefmark{1}Andrey~Lyakhov\\
	}
	
	\IEEEauthorblockA{\IEEEauthorrefmark{1}Institute for Information Transmission Problems, Russian Academy of Sciences, Moscow, Russia\\
		\IEEEauthorrefmark{2}National Research University Higher School of Economics, Moscow, Russia \\
		\IEEEauthorrefmark{3}Moscow Institute of Physics and Technology, Moscow, Russia \\
		Email: avdotin@wireless.iitp.ru, bankov@iitp.ru, e@khorov.ru, lyakhov@iitp.ru\\
	}
}

\maketitle
\copyrightnotice
\begin{abstract}
	In 2019 IEEE 802 LAN/MAN Standards Committee started the development of the next major amendment of the Wi-Fi standard: the IEEE 802.11be, also known as Wi-Fi 7.
	This new amendment will introduce many new functions and will improve the existing ones that will make Wi-Fi more efficient in many new scenarios.
	One of the scenarios is the service of Real-Time Applications with strict requirements on latency and reliability of communications.
	Providing low latencies can be challenging in Wi-Fi because of the unlicensed spectrum and related interference from neighboring devices.
	In this paper, we consider the usage of OFDMA transmissions for Real-Time Applications and design resource allocation algorithms that can provide the required latency and reliability in the presence of interference.
\end{abstract}

\section{Introduction}{\label{sec:intro}}
Currently, wireless networks are being developed in three main directions: higher throughputs, bigger numbers of devices served in the Internet of Things scenarios, and lower latency in time-sensitive applications.
The last direction has attracted the attention of the IEEE 802.11 Working Group (WG), which in 2019 created a Real-Time Applications (RTA) Topic Interest Group (TIG).
The RTA TIG has developed a list of scenarios of interest, which include virtual and augmented reality, online gaming, remote control, and industrial automation.
These scenarios are related to applications that are sensitive to delay, which shall be lower than $1$--\SI{10}{\ms}, and to the packet loss rate, which shall not exceed $10^{-8}$--$10^{-5}$ \cite{discussion_target_presentation,usecases_presentation}.
The RTA TIG has proposed many potential solutions \cite{rta_tig} to enable Wi-Fi networks to serve such strict applications. These solutions will be studied and refined during the development of the next main amendment of the Wi-Fi standard: the 802.11be \cite{be_par}, also known as Extremely High Throughput (EHT).

802.11be will introduce many new PHY and MAC layer solutions and will develop the solutions proposed in previous amendments like 802.11ac and 802.11ax.
802.11be will increase the bandwidth to \SI{320}{\MHz}, the number of MIMO spatial streams to 16, will provide solutions for the non-continuous spectrum usage, and will likely add completely new features, such as hybrid adaptive automatic repeat request (HARQ) and coordinated and joint transmissions of multiple access points (APs).
The exhaustive list of the features currently discussed in the context of 802.11be is described in~\cite{wifi7}.  

One of the features that will be improved in 802.11be is the orthogonal frequency division multiple access (OFDMA).
It has been introduced in 802.11ax both for downlink and uplink transmissions and is considered as a potential enabler of RTA in Wi-Fi.
Currently, with OFDMA, an AP can divide the frequency band into resource units (RUs) and allocate them either for deterministic transmissions by specific stations (STAs) or for random access (RA).
The latter is useful for uplink transmissions in cases when the AP does not know which STA currently needs channel resources and allows a STA to use shared channel resources to transmit a data frame or to request channel resources for deterministic transmissions.

Purely deterministic transmissions with OFDMA might be an inappropriate solution for RTA when STAs generate irregular uplink traffic, because requesting the channel resources introduces additional latency.
However, the AP can combine the joint usage of random and deterministic transmissions with OFDMA to organize implicit signaling about the STAs' need for channel resources \cite{avdotin2019ofdma}.
We have previously shown \cite{avdotin2019enabling} that such an approach can be further extended in 802.11be if the new amendment improves the OFDMA procedure in Wi-Fi.
Specifically, in 802.11ax one RU cannot be allocated for deterministic transmission by several STAs, but if 802.11be allows such an action, the AP will be able to distribute the channel resources more uniformly between the STAs and thus decrease the probability of collisions that lead to packet losses.

Currently, the existing algorithms for OFDMA resource allocation have been designed for scenarios when transmission attempts can fail only due to the collisions, but Wi-Fi networks can also suffer from the channel noise or interference from neighboring devices.
In this paper, we show that the existing solutions operate poorly in the case of the non-ideal channel and design new algorithms that can provide reliable service for RTA traffic even in the presence of noise or interference.

The rest of the paper is organized in the following way.
Section~\ref{sec:ofdma} provides the basics of OFDMA in Wi-Fi networks.
Section~\ref{sec:litra} reviews the prior arts on RTA in Wi-Fi.
Section~\ref{sec:problem} introduces the studied scenario and states the problem.
In Section~\ref{sec:algorithms}, we describe the developed new algorithms for OFDMA resource allocation in Wi-Fi.
In Section~\ref{sec:results}, we evaluate the efficiency of the developed algorithms.
Section~\ref{sec:conclusion} contains the conclusion.

\section{OFDMA in Wi-Fi}
\label{sec:ofdma}
\begin{figure}[h!]
	\centering
	\begin{tikzpicture}[scale = 1.0]
	\footnotesize
	\draw [arrows={-triangle 45}] (0.3,0.8) -- (7.5,0.8);
	\draw [arrows={-triangle 45}] (0.3,0.8) -- (0.3,3.2);
	\node at (7.3,  0.5) {\textit{Time}};
	\node [rotate=90] at (0,  1.8) {\textit{Frequency}};
	\draw [line width=0.5mm] (1, 0.8) rectangle (2, 2.6);
	\node [text width=1.5cm, align=center] at (1.5,  1.8) {TF};
	\draw [line width=0.5mm] (3, 0.8) rectangle (5, 1.3);
	\draw [line width=0.5mm] (3, 1.3) rectangle (5, 1.8);
	\draw [line width=0.5mm] (3, 1.8) rectangle (5, 2.1);
	\draw [line width=0.5mm] (3, 2.1) rectangle (5, 2.6);
	\node [text width=2cm, align=center] at (4, 1.05) {RU 1};
	\node [text width=2cm, align=center] at (4, 1.55) {RU 2};
	\node [text width=2cm, align=center] at (3.85, 1.95) {\quad...};
	\node [text width=2cm, align=center] at (4, 2.35) {RU $F$};
	\draw [line width=0.5mm] (6, 0.8) rectangle (7.2, 2.6);
	\node [text width=1.5cm, align=center] at (6.6,  1.8) {MSBA};
	\draw (2.0,  0.3) -- (2.0,  0.8);
	\draw (3.0,  0.3) -- (3.0,  0.8);
	\draw [arrows={triangle 45-triangle 45}] (2.0,0.3) -- (3.0,0.3);
	\node at (2.5,  0.5) {\small$SIFS$};
	\draw (5.0,  0.3) -- (5.0,  0.8);
	\draw (6.0,  0.3) -- (6.0,  0.8);
	\draw [arrows={triangle 45-triangle 45}] (5.0,0.3) -- (6.0,0.3);
	\node at (5.5,  0.5) {\small$SIFS$};
\end{tikzpicture}
	\vspace{-1em}
	\caption{The uplink OFDMA sequence}
	\label{fig:ofdma}
\end{figure}
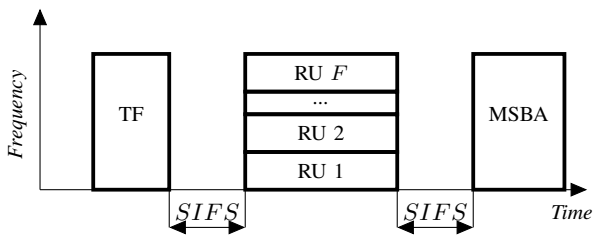

OFDMA in Wi-Fi has been introduced in the 802.11ax standard amendment for downlink and uplink transmissions and is used together with legacy Wi-Fi channel access.
Let us describe a typical uplink transmission procedure (see Fig. \ref{fig:ofdma}) with OFDMA since, in this paper, we focus on uplink optimization for RTA.
The OFDMA transmissions are controlled by the AP and start with a trigger frame (TF) broadcasted by the AP.
The purpose of the TF is to synchronize the transmissions by many STAs, to deliver the information about the scheduled channel resources, and to inform the surrounding STAs that the channel will be busy during the OFDMA transmission.
The OFDMA transmission by the STAs starts Short Inter-Frame Spacing (SIFS) after the TF.
SIFS after the OFDMA transmissions by the STAs, the AP sends a Multi-Station Block Acknowledgement (MSBA) frame to acknowledge the successful transmissions.

The AP divides the available channel resources into resource units (RU), which define the time-frequency block that can be used for STA transmissions.
Within a single OFDMA transmission, all the allocated RUs have the same duration, which has an upper limit of $\approx \SI{5}{\ms}$ and is determined by the longest transmission.
The frequency width of RUs allocated for different STAs can be different: the smallest RU has a width of 26 OFDM tones which can be aggregated into 52-, 106-, 242-, 484-, and 996-tone RUs (the numbers are not all divisible by 26 because additional guard tones are inserted).
The aggregation of RUs has some limitations related to the simplicity considerations, and the placing of pilot tones, e.g., 26-tone RUs \#1 and \#2 can be aggregated into a 52-tone RU, but RUs \# 2 and \# 3 cannot.
The details are described in \cite{khorov2018tutorial}.
The \SI{20}{\MHz} band corresponds to a 242-tone RU, so OFDMA allows serving 9 STAs at a time.

The RUs can be allocated either for deterministic transmission by specific STAs, or for all the STAs at once to transmit with Uplink OFDMA Random Access (UORA).
In 802.11ax, a single STA can use no more than one RU (but the width of this RU can vary), and several STAs cannot use the same RU for deterministic transmissions.
The situation is different for UORA, which works as follows.
A STA generates a backoff counter $r$ randomly from the interval $\left[0, OCW_{min}\right]$, where $OCW_{min}$ is the contention window.
After receiving every TF, the STA compares $r$ with the number of RUs allocated for UORA and, if $r$ is lower than this number, it transmits its data in a randomly chosen RU.
If this transmission is unsuccessful, e.g., because of a collision caused by some other STA's transmission in the same RU, the STA doubles its contention window and repeats the procedure.
The minimal and maximal contention windows are the parameters specified by the AP.

In the new 802.11be amendment, some restrictions on OFDMA introduced in 802.11ax, such as the limit on one RU allocated to one STA, may be removed.
In \cite{avdotin2019enabling}, we show that allowing several STAs to transmit in the same RU in case of non-saturated traffic can help to equalize the load between many RUs and thus to decrease the collision probability.
In this paper, we show that allowing a STA to use several RUs can also be an efficient solution to cope with noise and interference from hidden STAs.

\section{Related Works}
\label{sec:litra}
Solutions for RTA in Wi-Fi are actively proposed both in the IEEE 802.11 WG and in scientific papers.

Some solutions are rather straightforward, e.g., to introduce a special enhanced distributed channel access (EDCA) access category for time-sensitive traffic \cite{genc2019wi}.
A variant of such a solution is used in the Auto Repetition (AR) mode designed to service low latency flows.
With AR, a STA repeatedly transmits a packet according to some special rules in order to achieve ``zero'' PER in the 802.11 links.
AR requires an additional EDCA queue for the low latency packets, which would have its own backoff counter, which should be should be counted down in a way to enable frame transmission for a given number of times within the latency budget.

An example of such solutions is the usage of multiple links on adjacent physical channels within a band with one PHY \cite{rodriguezmulti}.
The authors propose for each STA to run separate carrier sense functions on different channels and transmit data on one link at a time.
Such a solution can significantly decrease the channel access delay, which is useful for RTA.

Another solution \cite{adhikariproposals} is to semi-persistently allocate narrow bandwidth resources for expedited transmission of low latency data. 
Also, it will be useful to enable the sharing of TXOP between Non-AP STAs and their AP and sharing of TXOPs between nodes across multiple BSSs via inter-BSS coordination.
These solutions can reduce the number of channel access attempts and thus decrease delays.

A possible improvement of OFDMA for RTA is the propose the Uplink Persistent Allocation (UPA) \cite{zuouse}.
This mechanism has been proposed for 802.11be to allow the AP to schedule RUs to a STA for a long time.
For that, the AP sends a UPA announcement frame which establishes the channel assignment to a STA.
Afterward, the AP can transmit short TF, which instead of long user info field for each user contains a short UPA indication (1 octet per user instead of 6), thus saving channel resources.

In \cite{avdotin2019ofdma}, we have developed a Cyclic Resource Allocation (CRA) algorithm to allocate OFDMA resources in the 802.11ax network in order to provide low latency with low packet loss ratio (PLR).
This algorithm uses the UORA functionality to detect situations when many STAs require channel resources. When such a situation is detected, it assigns RUs to all RTA STAs one by one.
This solution ensures that in case of necessity, every STA receives an RU, but at the same time, it does not use too many channel resources when the STAs do not need resources.
We have proposed a better algorithm, the Group Resource Allocation (GRA) in \cite{avdotin2019enabling} that requires modifications of the standard OFDMA procedure, which will be possible in the new 802.11be amendment.
When the AP detects a situation when a STA requires resources, it divides the STAs into equal groups and assigns an RU to each group.
We have shown that this algorithm performs better than CRA when the STAs generate non-saturated traffic, which is a typical scenario for RTA.
The CRA and GRA algorithms have a drawback: when assigning RUs to the STAs they do not take into account the possibility of STA data to be lost due to noise or due to the interference from non-OFDMA STAs.
In this paper, we show that these algorithms perform poorly in the case of a noisy channel and design more robust resource allocation algorithms for RTA.

\section{Problem Statement}
\label{sec:problem}
We consider a Wi-Fi network with one AP and $N$ RTA STAs.
The STAs generate a Poisson flow of data packets with intensity $\lambda$, which should be delivered to the AP within a delay budget $D$.
The packets which are not transmitted within the delay budget are lost.
The data packets generated by all STAs have the same length and can be transmitted with a single transmission in a 26-tone RU.

The STAs transmit their data with OFDMA using the RUs scheduled by the AP.
The period of TF transmission (further called \emph{slot}) by the AP equals $T$.
Let the channel width be equal to $F$ 26-tone RUs, out of which a maximum of $k$ 26-tone RUs can be allocated for the RTA STAs.
The STAs use the assigned RUs, which can be allocated either for deterministic transmissions or RA.
If more than one STA tries to transmit a frame in one RU at a time, then a collision happens, and the transmission attempt is unsuccessful.
We also assume that the channel is non-ideal. In essence, even if only one STA tries to transmit a frame in an RU, the transmission can be unsuccessful with probability $p$ due to the random channel noise or due to the interference from hidden STAs.

For such a scenario, we state the problem \emph{to develop an RU scheduling algorithm that can provide the packet loss rate less than $PLR_{max}$ and consume as little channel resources as possible}.

\section{Resource Assignment Algorithms for RTA}
\label{sec:algorithms}
In this section, we develop resource allocation algorithms aimed at minimization of RTA data transmission delay.
Since all these algorithms aim at minimizing delay, we assume that the network uses the minimal and maximal contention window set to zero to let the AP resolve the collisions instead of the STAs.

The main problem that the AP faces while scheduling RUs is that it does not know a priori which STAs currently need channel resources.
To solve this problem, the AP can wait for the STAs to transmit buffer status reports, but their transmission introduces additional delay comparable with the delay budget of several milliseconds.
For such a reason, we propose using an implicit way to determine the traffic.
For all the further presented algorithms, it works as follows.

Initially, the AP operates in the \emph{waiting mode}.
In this mode, it allocates a single RU for RA.
When any STA generates a data frame, it tries to transmit the frame in the allocated RU.
If this STA is the only STA which currently has data to transmit, its transmission will be successful.
Otherwise, there will be a collision in this RU, which is a signal that some STAs have uplink packets for transmission.
The AP should allocate more RUs but does not know to which STAs exactly.
In such a situation, the AP switches to the \emph{collision resolution mode}.

We further present three algorithms which differ in the AP behavior during the collision resolution mode.

\subsection{Noise Resistant Uplink OFDMA Random Access (NUORA)}
The first algorithm is an extension of UORA for noisy channels.
With standard UORA, when a STA makes an unsuccessful transmission, it makes a new transmission attempt in the next slot.
Such repetitions increase the delay.
At the same time, repeating the frame in the frequency direction instead of time does not increase the delay.
With NUORA, the AP allocates $k$ RUs only for RA, and the STA, which has data, sends a copy of its frame in $f$ randomly-chosen RUs, where $f$ is the algorithm parameter.

The AP allocates $k$ RUs for RA as long as there are unsuccessful transmissions in the allocated RUs.
If there are only empty RUs or RUs with successful transmissions during the slot, the AP switches back to the waiting mode and allocates one RU for RA.

\begin{figure}[h!]
	\centering
	\begin{tikzpicture}[scale=0.6]
    \footnotesize
    \draw [arrows={-triangle 45}] (0,1) -- (13,1);
    \node [text width=2cm, align=center] at (12.5,0.6) {\textit{Time}};
    \draw [arrows={-triangle 45}] (0.3,1) -- (0.3,5.0);
    \node [text width=2cm, rotate=90, align=center] at (0,2.5) {\textit{Frequency}};
    \draw [pattern=north west lines, pattern color=red] [line width=0.5mm] (0.5,1) rectangle (3.5,2);
    \node [text width=2cm, align=left] at (2.3,1.5) {RA: 1, 2, 3};

    \draw [line width=0.5mm] (3.5,1) rectangle (6.5,2);
    \node [text width=2cm, align=left] at (5.3,1.5) {RA: ---};
    \draw [pattern=north west lines, pattern color=red] [line width=0.5mm] (3.5,2) rectangle (6.5,3);
    \node [text width=2cm, align=left] at (5.3,2.5) {RA: 1, 2, 3};
    \draw [pattern=north west lines, pattern color=red] [line width=0.5mm] (3.5,3) rectangle (6.5,4);
    \node [text width=2cm, align=left] at (5.3,3.5) {RA: 1, 3};
    \draw [pattern=north east lines, pattern color=green] [line width=0.5mm] (3.5,4) rectangle (6.5,5);
    \node [text width=2cm, align=left] at (5.3,4.5) {RA: 2};

    \draw [pattern=north east lines, pattern color=green] [line width=0.5mm] (6.5,1) rectangle (9.5,2);
    \node [text width=2cm, align=left] at (8.5,1.5) {RA: 3};
    \draw [pattern=north west lines, pattern color=red] [line width=0.5mm] (6.5,2) rectangle (9.5,3);
    \node [text width=2cm, align=left] at (8.5,2.5) {RA: 1, 3};
    \draw [line width=0.5mm] (6.5,3) rectangle (9.5,4);
    \node [text width=2cm, align=left] at (8.5,3.5) {RA: ---};
    \draw [pattern=north east lines, pattern color=green] [line width=0.5mm] (6.5,4) rectangle (9.5,5);
    \node [text width=2cm, align=left] at (8.5,4.5) {RA: 1};

    \draw [line width=0.5mm] (9.5,1) rectangle (12.5,2);
    \node [text width=2cm, align=left] at (11.5,1.5) {RA: ---};

%    \draw [line width=0.5mm] (3.0,2) rectangle (5.5,3);
%    \node [text width=2cm, align=center] at (4,2.2) {RA};
%    \node [text width=2cm, align=center] at (3,2.2) {$1$};
%    \draw [pattern=north west lines, pattern color=red] [line width=0.5mm] (3,3) rectangle (5,5);
%    \node [text width=2cm, align=center] at (4,3) {RA};
%    \node [text width=2cm, align=center] at (3,3) {$2$};
%    \draw [pattern=north east lines, pattern color=green] [line width=0.5mm] (3.0,3.4) rectangle (5.5,4.2);
%    \node [text width=2cm, align=center] at (4,3.8) {RA};
%    \node [text width=2cm, align=center] at (3,3.8) {$3$};
%    
%    \draw [pattern=north east lines, pattern color=green] [line width=0.5mm] (4.5,1) rectangle (6.5,1.8);
%    \node [text width=2cm, align=center] at (6,1.4) {RA};
%    \node [text width=2cm, align=center] at (5,1.4) {$0$};
%    \draw [line width=0.5mm] (4.5,1.8) rectangle (6.5,2.6);
%    \node [text width=2cm, align=center] at (6,2.2) {RA};
%    \node [text width=2cm, align=center] at (5,2.2) {$1$};
%    \draw [pattern=north east lines, pattern color=green] [line width=0.5mm] (4.5,2.6) rectangle (6.5,3.4);
%    \node [text width=2cm, align=center] at (6,3) {RA};
%    \node [text width=2cm, align=center] at (5,3) {$2$};
%    \draw [line width=0.5mm] (4.5,3.4) rectangle (6.5,4.2);
%    \node [text width=2cm, align=center] at (6,3.8) {RA};
%    \node [text width=2cm, align=center] at (5,3.8) {$3$};
%
%    \draw [line width=0.5mm] (6.5,1) rectangle (8.5,1.8);
%    \node [text width=2cm, align=center] at (7.5,1.4) {RA};
\end{tikzpicture}
	\vspace{-1em}
	\caption{An example of NUORA operation with parameters $k = 4$, $f = 2$. STAs 1, 2, 3 have data to transmit.}
	\label{fig:uora}
\end{figure}
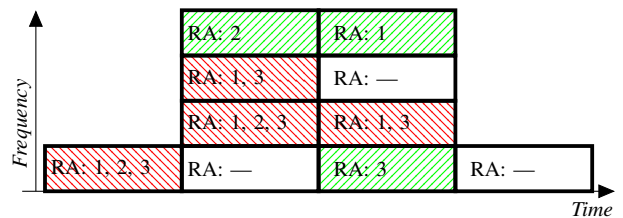

An example of NUORA operation is shown in Fig. \ref{fig:uora}, where red hatching shows collision RUs, green hatching shows successful RUs, and numbers mean the STAs that try to transmit in corresponding RUs.

Even though repeating the frames increases the load on the channel and thus the probability of collision, frame repetition can improve the reliability against the noise.

\subsection{Noise Resistant Group Resource Allocation (NGRA)}
The idea of NGRA is to spread the load over the RUs in a more uniform fashion, excluding the situation when most contending STAs choose the same RU, while most of the remaining RUs are left unused.
The GRA algorithm presented in \cite{avdotin2019enabling} solves this problem by dividing the STAs into equal-sized groups and allocating an RU to each group.
We extend this idea in NGRA to make it more robust against noise by making $f$ transmission attempts within one slot.

When the AP switches to the collision resolution mode, it marks all the STAs as requiring resources.
Then the AP for each marked STA assigns $f$ random RUs, but while assigning resources for each STA the AP excludes the RUs assigned to $\lceil\frac{fN}{k}\rceil$ STAs.
The STAs transmit $f$ copies of their data frames in the assigned RUs.
After the slot, the AP marks all the STAs which correspond to RUs that did not contain any unsuccessful transmissions as not requiring resources.
In the next slot, the AP allocates $k - 1$ RUs to the STAs, which are still marked using the procedure described above, and also allocates one RU for RA, which can be used by the unmarked STAs if they suddenly generate more data.
The AP repeats the described procedure, and if there are collisions in the RU allocated for RA, then all the STAs which were previously marked as not needing resources are marked as needing resources again.
When all the RTA STAs are not marked as requiring resources, the AP stops the procedure and switches back to the waiting mode and allocates one RU for RA.

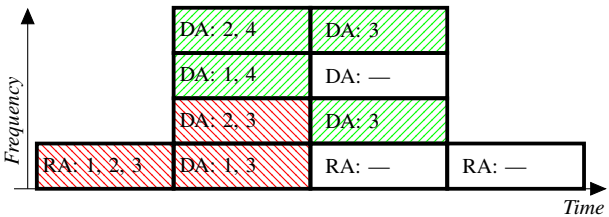
\begin{figure}[h!]
	\centering
	\begin{tikzpicture}[scale=0.6]
    \footnotesize
    \draw [arrows={-triangle 45}] (0,1) -- (13,1);
    \node [text width=2cm, align=center] at (12.5,0.6) {\textit{Time}};
    \draw [arrows={-triangle 45}] (0.3,1) -- (0.3,5.0);
    \node [text width=2cm, rotate=90, align=center] at (0,2.5) {\textit{Frequency}};
    \draw [pattern=north west lines, pattern color=red] [line width=0.5mm] (0.5,1) rectangle (3.5,2);
    \node [text width=2cm, align=left] at (2.3,1.5) {RA: 1, 2, 3};

    \draw [pattern=north west lines, pattern color=red] [line width=0.5mm] (3.5,1) rectangle (6.5,2);
    \node [text width=2cm, align=left] at (5.3,1.5) {DA: 1, 3};
    \draw [pattern=north west lines, pattern color=red] [line width=0.5mm] (3.5,2) rectangle (6.5,3);
    \node [text width=2cm, align=left] at (5.3,2.5) {DA: 2, 3};
    \draw [pattern=north east lines, pattern color=green] [line width=0.5mm] (3.5,3) rectangle (6.5,4);
    \node [text width=2cm, align=left] at (5.3,3.5) {DA: 1, 4};
    \draw [pattern=north east lines, pattern color=green] [line width=0.5mm] (3.5,4) rectangle (6.5,5);
    \node [text width=2cm, align=left] at (5.3,4.5) {DA: 2, 4};

    \draw [line width=0.5mm] (6.5,1) rectangle (9.5,2);
    \node [text width=2cm, align=left] at (8.5,1.5) {RA: ---};
    \draw [pattern=north east lines, pattern color=green] [line width=0.5mm] (6.5,2) rectangle (9.5,3);
    \node [text width=2cm, align=left] at (8.5,2.5) {DA: 3};
    \draw [line width=0.5mm] (6.5,3) rectangle (9.5,4);
    \node [text width=2cm, align=left] at (8.5,3.5) {DA: ---};
    \draw [pattern=north east lines, pattern color=green] [line width=0.5mm] (6.5,4) rectangle (9.5,5);
    \node [text width=2cm, align=left] at (8.5,4.5) {DA: 3};

    \draw [line width=0.5mm] (9.5,1) rectangle (12.5,2);
    \node [text width=2cm, align=left] at (11.5,1.5) {RA: ---};

%    \draw [line width=0.5mm] (3.0,2) rectangle (5.5,3);
%    \node [text width=2cm, align=center] at (4,2.2) {RA};
%    \node [text width=2cm, align=center] at (3,2.2) {$1$};
%    \draw [pattern=north west lines, pattern color=red] [line width=0.5mm] (3,3) rectangle (5,5);
%    \node [text width=2cm, align=center] at (4,3) {RA};
%    \node [text width=2cm, align=center] at (3,3) {$2$};
%    \draw [pattern=north east lines, pattern color=green] [line width=0.5mm] (3.0,3.4) rectangle (5.5,4.2);
%    \node [text width=2cm, align=center] at (4,3.8) {RA};
%    \node [text width=2cm, align=center] at (3,3.8) {$3$};
%    
%    \draw [pattern=north east lines, pattern color=green] [line width=0.5mm] (4.5,1) rectangle (6.5,1.8);
%    \node [text width=2cm, align=center] at (6,1.4) {RA};
%    \node [text width=2cm, align=center] at (5,1.4) {$0$};
%    \draw [line width=0.5mm] (4.5,1.8) rectangle (6.5,2.6);
%    \node [text width=2cm, align=center] at (6,2.2) {RA};
%    \node [text width=2cm, align=center] at (5,2.2) {$1$};
%    \draw [pattern=north east lines, pattern color=green] [line width=0.5mm] (4.5,2.6) rectangle (6.5,3.4);
%    \node [text width=2cm, align=center] at (6,3) {RA};
%    \node [text width=2cm, align=center] at (5,3) {$2$};
%    \draw [line width=0.5mm] (4.5,3.4) rectangle (6.5,4.2);
%    \node [text width=2cm, align=center] at (6,3.8) {RA};
%    \node [text width=2cm, align=center] at (5,3.8) {$3$};
%
%    \draw [line width=0.5mm] (6.5,1) rectangle (8.5,1.8);
%    \node [text width=2cm, align=center] at (7.5,1.4) {RA};
\end{tikzpicture}
	\vspace{-1em}
	\caption{An example of NGRA operation with parameters $N = 4$, $k = 4$, $f = 2$. STAs 1, 2, 3 have data to transmit.}
	\label{fig:ngra}
	
\end{figure}

An example of NGRA operation is shown in Fig. \ref{fig:ngra}, where ``DA: x, y'' means that the RU is assigned for deterministic transmissions by STAs x and y.

NGRA allows several STAs to use the same RU, which is a feasible solution in the case of low RTA traffic intensity.
However, if the traffic becomes intense, such a solution may suffer from collisions.

\subsection{Noise Resistant Cyclic Resource Allocation (NCRA)}
Another approach is the improved version of the CRA algorithm presented in \cite{avdotin2019enabling}.
With NCRA, the AP allocates one RU for RA, while the remaining RUs are assigned to no more than one STA cyclically, allowing each STA to transmit $f$ copies of their frames.
Specifically, the AP allocates RU 0 for RA, and then allocates RUs $1...f$ to STA 1, RUs $f+1...2f$ to STA 2, etc., until it reaches the last available RU.
The STAs that have been allocated RUs use these RUs to transmit their frames, while the remaining STAs can use the RU allocated for RA if they have any data.
After the slot, the AP examines the RUs, and assigns RUs again to the STAs, which could not transmit any copy of their data in the assigned RUs.
Then the AP keeps cycling through the remaining STAs, assigning $f$ RUs to each STA.
Such a procedure repeats as long as there are STAs that could not transmit any copy of their frames in the allocated RUs or there is an unsuccessful transmission in RA.
Afterward, the AP switches back to the waiting mode and allocates one RU for RA.

\begin{figure}[h!]
	\centering
	\begin{tikzpicture}[scale=0.6]
    \footnotesize
    \draw [arrows={-triangle 45}] (0,1) -- (13,1);
    \node [text width=2cm, align=center] at (12.5,0.6) {\textit{Time}};
    \draw [arrows={-triangle 45}] (0.3,1) -- (0.3,6.0);
    \node [text width=2cm, rotate=90, align=center] at (0,2.5) {\textit{Frequency}};
    \draw [pattern=north west lines, pattern color=red] [line width=0.5mm] (0.5,1) rectangle (3.5,2);
    \node [text width=2cm, align=left] at (2.3,1.5) {RA: 1, 2, 4, 5};

    \draw [pattern=north west lines, pattern color=red] [line width=0.5mm] (3.5,1) rectangle (6.5,2);
    \node [text width=2cm, align=left] at (5.3,1.5) {RA: 4, 5};
    \draw [pattern=north east lines, pattern color=green] [line width=0.5mm] (3.5,2) rectangle (6.5,3);
    \node [text width=2cm, align=left] at (5.3,2.5) {DA: 1};
    \draw [pattern=north east lines, pattern color=green] [line width=0.5mm] (3.5,3) rectangle (6.5,4);
    \node [text width=2cm, align=left] at (5.3,3.5) {DA: 1};
    \draw [pattern=north east lines, pattern color=green] [line width=0.5mm] (3.5,4) rectangle (6.5,5);
    \node [text width=2cm, align=left] at (5.3,4.5) {DA: 2};
    \draw [pattern=north east lines, pattern color=green] [line width=0.5mm] (3.5,5) rectangle (6.5,6);
    \node [text width=2cm, align=left] at (5.3,5.5) {DA: 2};

    \draw [pattern=north east lines, pattern color=green] [line width=0.5mm] (6.5,1) rectangle (9.5,2);
    \node [text width=2cm, align=left] at (8.5,1.5) {RA: 5};
    \draw [line width=0.5mm] (6.5,2) rectangle (9.5,3);
    \node [text width=2cm, align=left] at (8.5,2.5) {DA: 3};
    \draw [line width=0.5mm] (6.5,3) rectangle (9.5,4);
    \node [text width=2cm, align=left] at (8.5,3.5) {DA: 3};
    \draw [pattern=north east lines, pattern color=green] [line width=0.5mm] (6.5,4) rectangle (9.5,5);
    \node [text width=2cm, align=left] at (8.5,4.5) {DA: 4};
    \draw [pattern=north east lines, pattern color=green] [line width=0.5mm] (6.5,5) rectangle (9.5,6);
    \node [text width=2cm, align=left] at (8.5,5.5) {DA: 4};

    \draw [line width=0.5mm] (9.5,1) rectangle (12.5,2);
    \node [text width=2cm, align=left] at (11.5,1.5) {RA: ---};

%    \draw [line width=0.5mm] (3.0,2) rectangle (5.5,3);
%    \node [text width=2cm, align=center] at (4,2.2) {RA};
%    \node [text width=2cm, align=center] at (3,2.2) {$1$};
%    \draw [pattern=north west lines, pattern color=red] [line width=0.5mm] (3,3) rectangle (5,5);
%    \node [text width=2cm, align=center] at (4,3) {RA};
%    \node [text width=2cm, align=center] at (3,3) {$2$};
%    \draw [pattern=north east lines, pattern color=green] [line width=0.5mm] (3.0,3.4) rectangle (5.5,4.2);
%    \node [text width=2cm, align=center] at (4,3.8) {RA};
%    \node [text width=2cm, align=center] at (3,3.8) {$3$};
%    
%    \draw [pattern=north east lines, pattern color=green] [line width=0.5mm] (4.5,1) rectangle (6.5,1.8);
%    \node [text width=2cm, align=center] at (6,1.4) {RA};
%    \node [text width=2cm, align=center] at (5,1.4) {$0$};
%    \draw [line width=0.5mm] (4.5,1.8) rectangle (6.5,2.6);
%    \node [text width=2cm, align=center] at (6,2.2) {RA};
%    \node [text width=2cm, align=center] at (5,2.2) {$1$};
%    \draw [pattern=north east lines, pattern color=green] [line width=0.5mm] (4.5,2.6) rectangle (6.5,3.4);
%    \node [text width=2cm, align=center] at (6,3) {RA};
%    \node [text width=2cm, align=center] at (5,3) {$2$};
%    \draw [line width=0.5mm] (4.5,3.4) rectangle (6.5,4.2);
%    \node [text width=2cm, align=center] at (6,3.8) {RA};
%    \node [text width=2cm, align=center] at (5,3.8) {$3$};
%
%    \draw [line width=0.5mm] (6.5,1) rectangle (8.5,1.8);
%    \node [text width=2cm, align=center] at (7.5,1.4) {RA};
\end{tikzpicture}
	\vspace{-1em}
	\caption{An example of NCRA operation with parameters $N = 5$, $k = 5$, $f = 2$. STAs 1, 2, 4, 5 have data to transmit.}
	\label{fig:ax}
\end{figure}
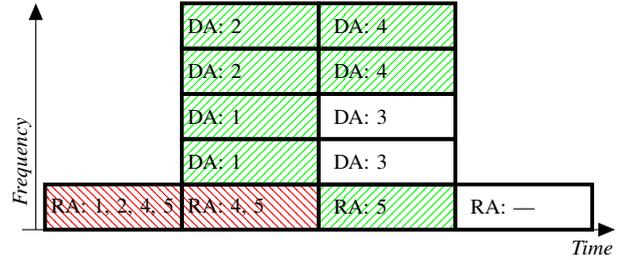        

An example of the NCRA operation in the case of a noiseless channel is shown in Fig. \ref{fig:ngra}.

Such an approach can be more efficient than NGRA in the case when the number of STAs is not very high.
At the same time, cycling through all the available STAs introduces additional delay because a STA has to wait for its turn to transmit.

\section{Numerical Results}
\label{sec:results}
\begin{figure}[htb]
	\centering
	\subfloat[$p = 0$]{\includegraphics[width=0.39\textwidth]{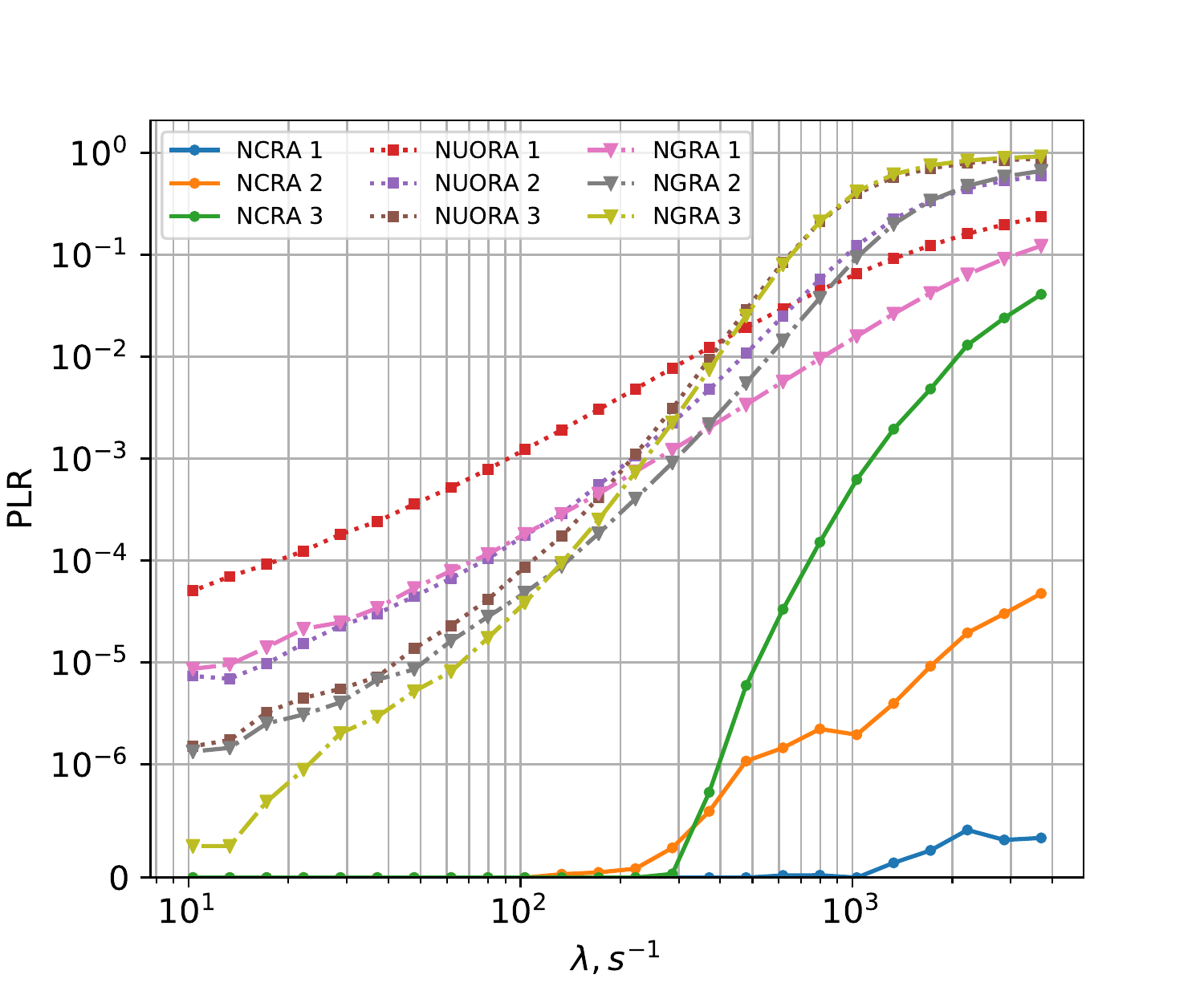}\label{fig:plr0}}\\        
	\subfloat[$p = 0.1$]{\includegraphics[width=0.39\textwidth]{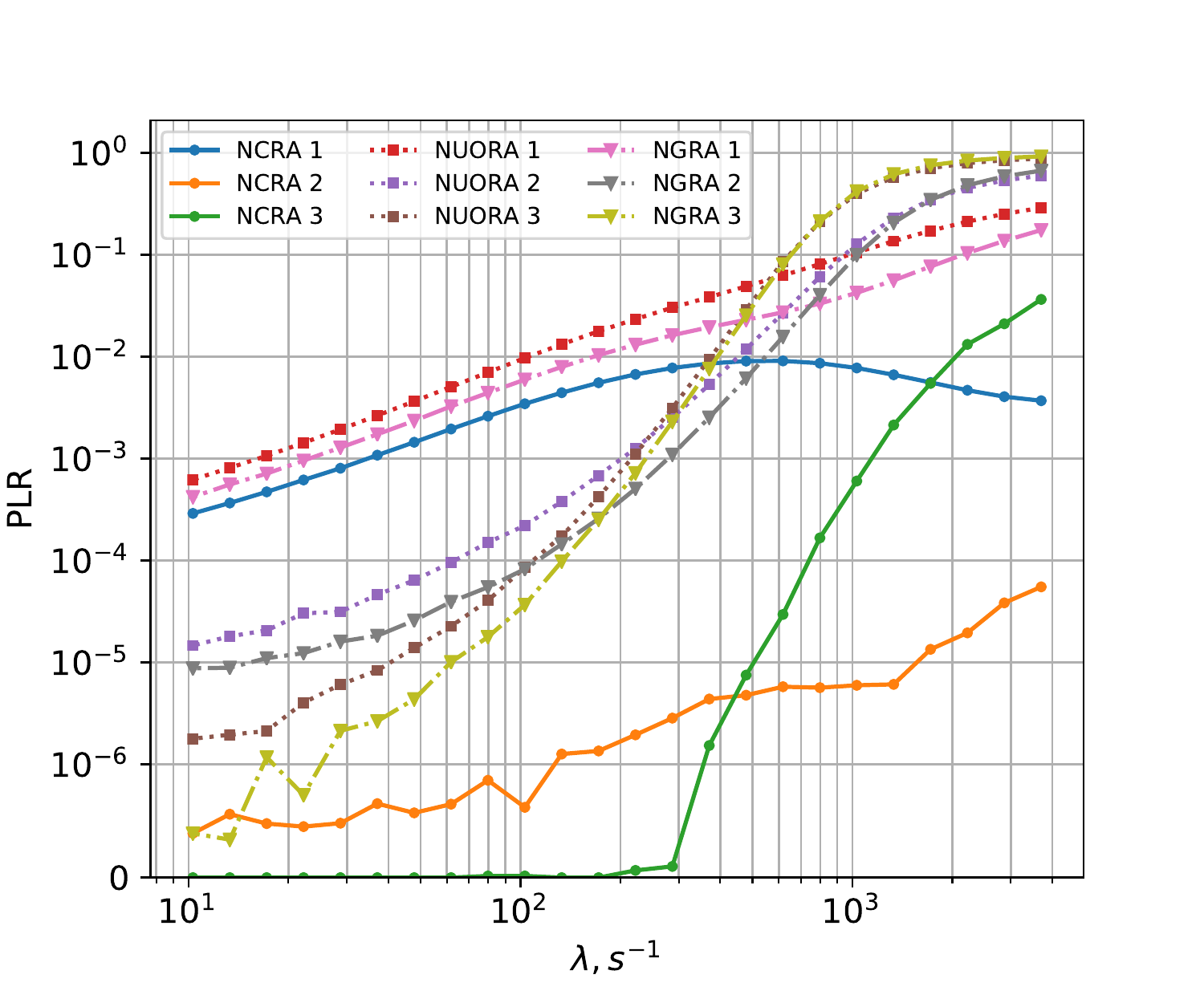}\label{fig:plr1}}\\
	\subfloat[$p = 0.2$]{\includegraphics[width=0.39\textwidth]{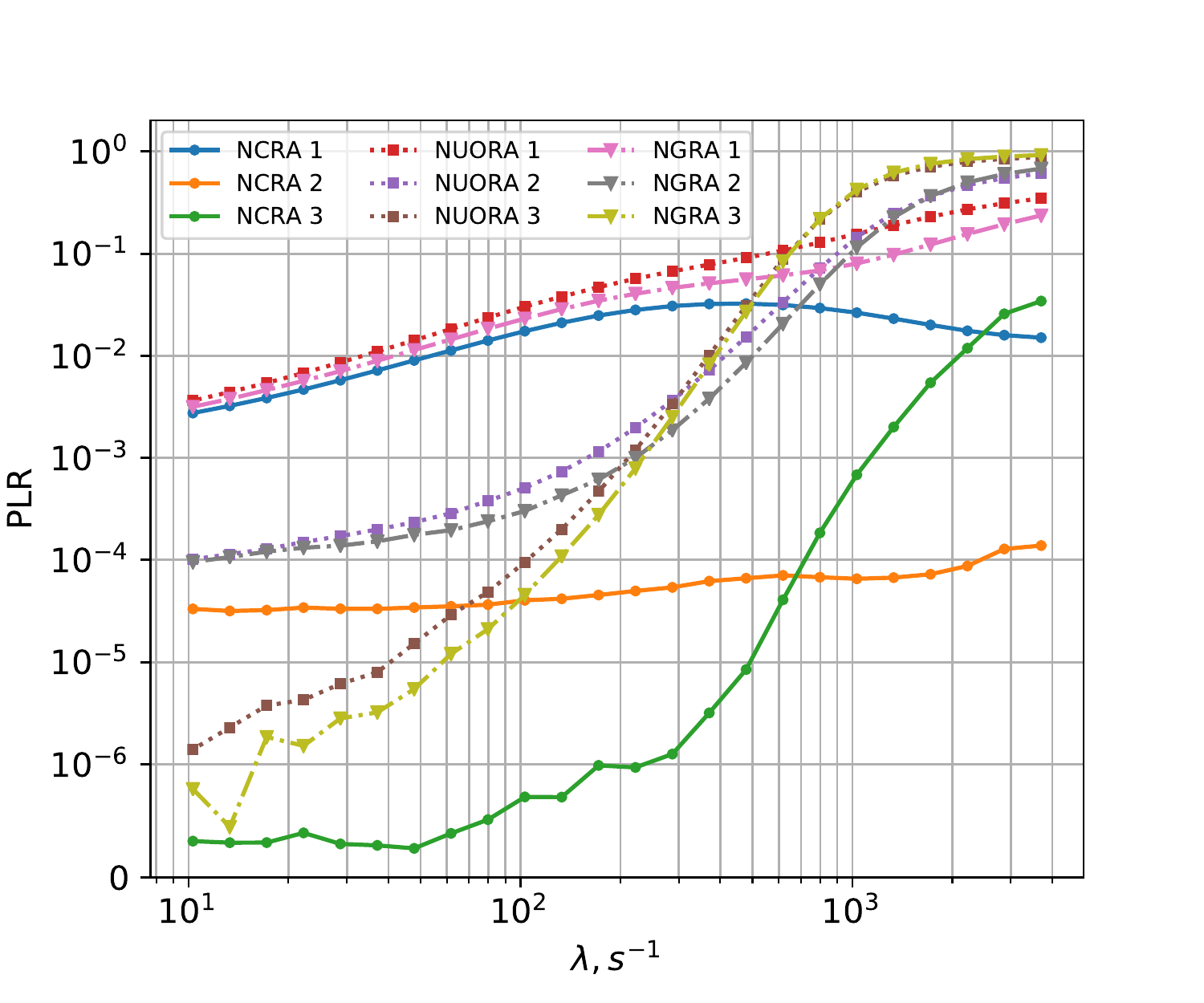}\label{fig:plr2}}
	\caption{The dependency of PLR on the packets arrival rate}
	\label{fig:plr}
	\vspace{-1em}
\end{figure}
\begin{figure}[htb]
	\centering
	\subfloat[$p = 0$]{\includegraphics[width=0.39\textwidth]{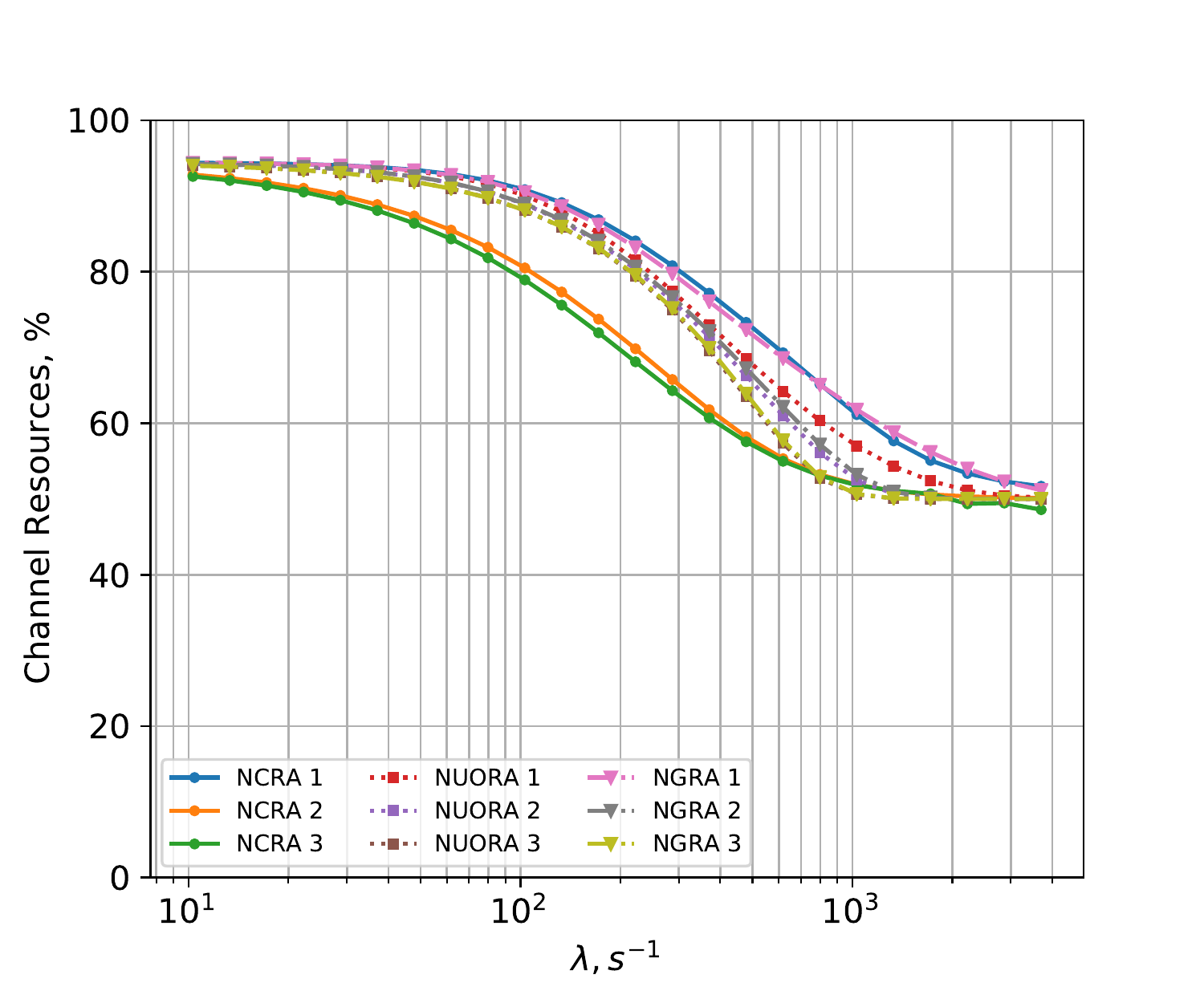}\label{fig:goodput0}}\\
	
	\subfloat[$p = 0.1$]{\includegraphics[width=0.39\textwidth]{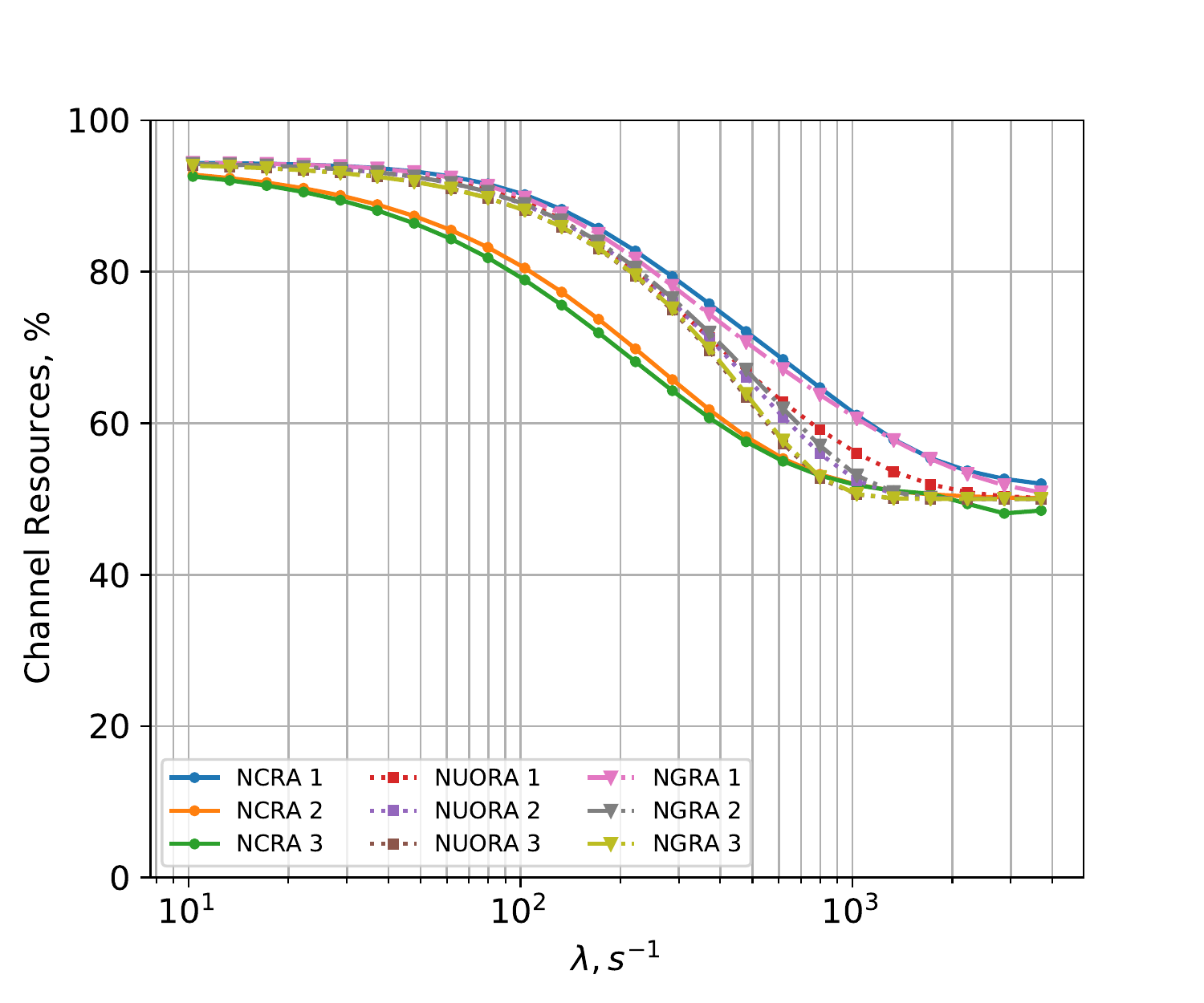}\label{fig:goodput1}}\\
	\subfloat[$p = 0.2$]{\includegraphics[width=0.39\textwidth]{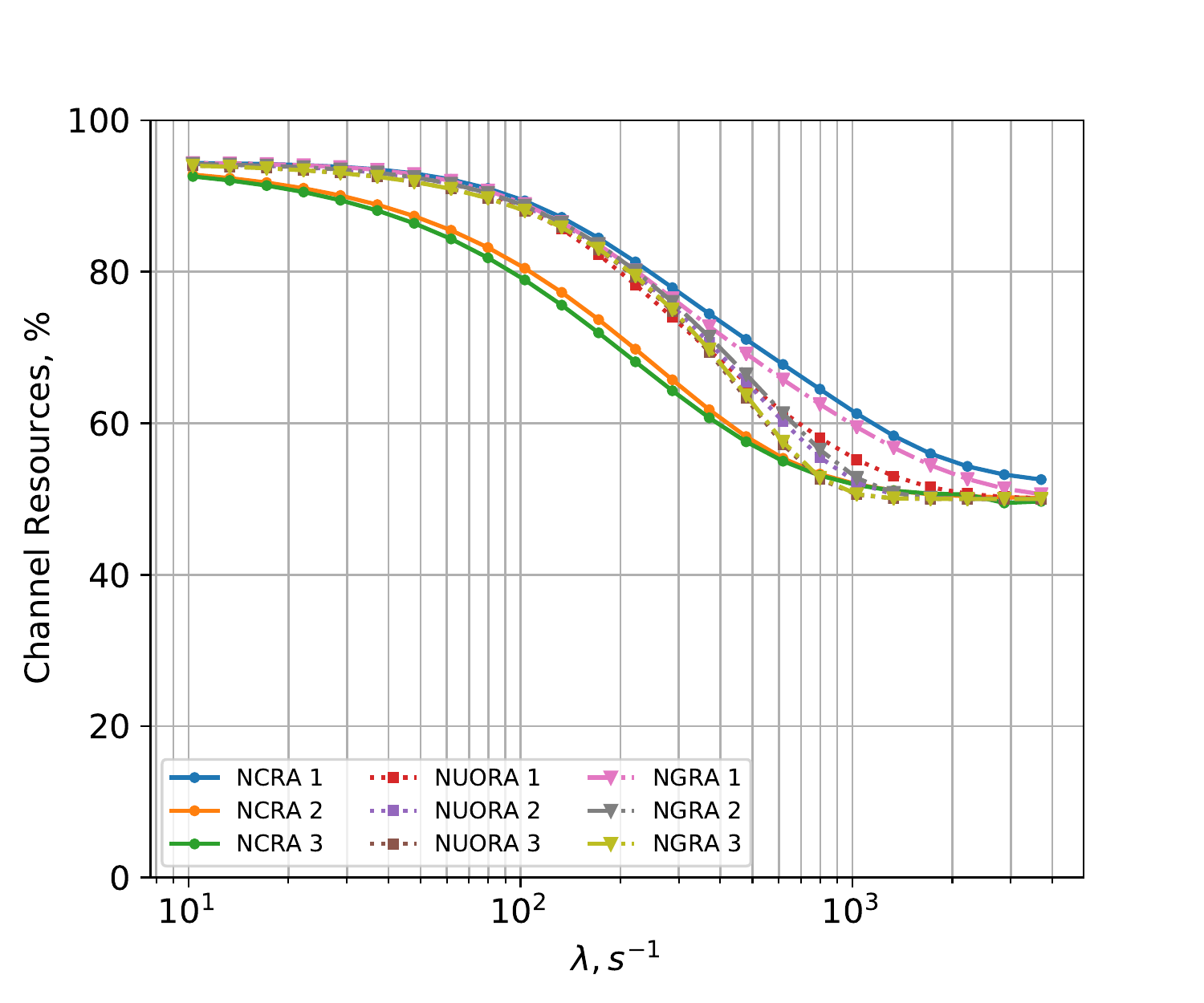}\label{fig:goodput2}}
	\caption{The dependency of the portion of channel resources available for Non-RTA STAs on the packets arrival rate}
	\label{fig:goodput}
	\vspace{-1em}
\end{figure}

To evaluate the performance of designed algorithms, we simulate the operation of a Wi-Fi network in the scenario described in Section \ref{sec:problem}.
We consider that there are $N = 18$ STAs in the network, and the network uses a \SI{40}{\MHz} channel, so the maximal number of 26-tone RUs is $F = 18$, out of which a maximum of $k = 9$ RUs can be allocated for RTA STAs.
The OFDMA slot duration equals $T \approx \SI{270}{\us}$, and the delay budget is $D = 5 T$, which corresponds to strict RTA scenario.
RTA STAs transmit short packets so that the size of 26-tone RUs is sufficient to transmit a packet at once.

We vary the RTA traffic intensity $\lambda$ and the probability $p$ of transmission in RU to be damaged by interference or noise.
We measure the PLR (see Fig. \ref{fig:plr}), i.e., the portion of packets discarded due to exceeding the delay budget, and the portion of channel resources available for Non-RTA STA transmissions (see Fig. \ref{fig:goodput}).
We consider the algorithms NUORA, NGRA, and NCRA, described in Section \ref{sec:algorithms} and try different values of parameter $f$: the number of frame copies that a STA makes during the contention resolution.
We show only the results for $f = 1, 2, 3$, because $f \geq 4 $ provides higher PLR for the considered range of traffic intensity.
When the parameter $f$ equals $1$, the algorithms NUORA, NGRA, and NCRA become the same as the algorithms UORA, GRA, and CRA presented and studied in \cite{avdotin2019enabling}.

In all scenarios, we see that with equal parameters, NGRA achieves lower PLR than NUORA.
Such a result is explained by the fact that NGRA groups the STAs and thus limits the collision probability.
At the same time, NGRA leaves a greater portion of channel resources for Non-RTA STAs, which means that in case of collisions, NGRA resolves them faster than NUORA.
In the considered scenario, NCRA reaches the lowest PLR because it excludes the collisions during the contention resolution phase.
However, such efficiency comes at the cost of a higher channel resource consumption, because the contention resolution phase for NCRA lasts much longer than for NGRA and NUORA.

In the scenario without noise shown in Figure \ref{fig:plr0}, increasing the number of frame repetitions $f$ for NGRA and NUORA can either decrease or increase the PLR depending on the traffic intensity.
If the traffic rate is not very high, making several transmission attempts in a slot increases the success probability: even if one copy of a frame is destroyed in a collision, another copy can still be delivered.
As a result, with $f = 3$ and low rate traffic NUORA and NGRA can achieve the PLR equal to $10^{-5}$: a typical requirement of RTA scenarios.
At the same time, with higher traffic rates, when more STAs are trying to send their data at a time, increasing the number of repetitions, on the contrary, increases the collision rate and thus the PLR.

In the case of NCRA, in a scenario without noise increasing, the parameter $k$ only increases the PLR and the channel resource consumption, because the repetition of frames increases the time it takes for the AP to cycle through all the STAs.

The situation is different in the case of the noisy channel (or the presence of hidden STAs).
For all the algorithms and the possibility that the frame will be lost due to the noise significantly increases the PLR, and for $f = 1$, none of the algorithms can satisfy the PLR requirement of $10^{-5}$.
Thus the previously presented algorithms CRA and GRA become inefficient for noisy channels.
At the same time, frame repetitions make the algorithms more stable against the noise.
Moreover, the results for NUORA and NGRA $f = 3$ almost do not change with increase of $p$, and even for $p = 0.2$ they can provide PLR lower than $10^{-5}$ for traffic intensity below~$\approx 6$ frames per second.

The results for NCRA are more sensitive to the noise, and generally we can conclure that the best number of repetitions depends on the intensity of noise: for $p = 0.1$, see Fig. \ref{fig:plr1}, it is better to use $f = 2$ repetitions, while for $p = 0.2$, see Fig. \ref{fig:plr2}, it is better to use $f = 3$ repetitions, but for very high traffic rate the PLR of $10^{-5}$ becomes unachievable.

As for the portion of channel resources left for Non-RTA STAs, it does not significantly depend on $p$ for all the algorithms.
It is almost the same for NUORA and NGRA, and it is lower for NCRA.
For NCRA, we can note the relation between the number of consumed channel resources with the PLR: the portion of remaining channel resources approaches to 50\% when the PLR is near $10^{-5}$.

\section{Conclusion}
\label{sec:conclusion}
In this paper, we have studied the problem of RTA service in new generation Wi-Fi networks with the usage of OFDMA.
We consider the scenario when RTA STAs transmit time-sensitive data in the uplink, and the AP solves the problem of OFDMA resource allocation to provide a low delay of $1$--$10$ milliseconds with a probability of packet loss not higher than $10^{-5}$.
The main problem that the AP faces while scheduling the resources is that it does not know which STAs require resources at the moment, and collecting the requests for channel resources from the STAs introduces a delay close to the delay budget of data frames.

To solve this problem, we propose using algorithms based on random access, which use implicit signaling about the requirements for channel resources from the STAs.
Such algorithms have been considered previously in \cite{avdotin2019ofdma, avdotin2019enabling}, but we show that these algorithms perform poorly in non-ideal channels when transmission attempts made by STAs can fail due to the random noise or interference from the hidden STAs.
We significantly extend these algorithms to cope with noise and interference: we propose for each STA to transmit several copies of their frames during one OFDMA transmission in different RUs.
We show that such an approach not only allows providing the low PLR for scenarios with high noise but can significantly reduce the PLR even for noiseless scenarios.

We have proposed three algorithms that implement this approach: NUORA, NGRA, and NCRA, out of which NGRA and NCRA show the best performance.
The algorithms have a configurable parameter $f$: the number of copies of the frame that the STA transmits during an OFDMA transmission.
We have shown that the best value of this parameter depends on the traffic rate and on the probability of frame loss due to noise or interference.
NCRA provides lower PLR than NGRA but consumes more channel resources, so the choice of which algorithm to use should be made according to the scenario requirements: if the traffic is low or the requirement on PLR is not too strict it is better to use NGRA, while for heavy traffic it is better to use NCRA.

As a direction of future work, we plan to develop an adaptive approach to configure the parameters of the proposed algorithms automatically.

\bibliographystyle{IEEEtran}
\bibliography{biblio.bib}

\end{document}